\documentclass[twocolumn,floatfix,pra,aps,showpacs,superscriptaddress]{revtex4}
\usepackage{epsfig,graphicx,tabularx}
\usepackage{amsmath,amsfonts,amssymb}
\usepackage{color,braket}

\newcommand{\be}{\begin{equation}}
\newcommand{\ee}{\end{equation}}

\begin{document}

\title{Magnetic phase transition in coherently coupled Bose gases in optical lattices}

\author{L. Barbiero}
\affiliation{Dipartimento di Fisica e Astronomia "Galileo Galiei", Universit\`a di Padova, 35131 Padova, Italy}
\author{M. Abad}
\affiliation{INO-CNR BEC Center and Dipartimento di Fisica, Universit\`a di Trento, 38123 Povo, Italy}
\affiliation{Quantum Systems Unit, OIST Graduate University, Onna, Okinawa 904-0495, Japan}
\author{A. Recati}
\affiliation{INO-CNR BEC Center and Dipartimento di Fisica, Universit\`a di Trento, 38123 Povo, Italy}
\affiliation{Technische Universit\"at M\"unchen, James-Franck-Stra{\ss}e 1, 85748 Garching, Germany}

\begin{abstract}
We describe the ground state of a gas of bosonic atoms with two coherently coupled internal levels in a deep optical lattice in a one dimensional geometry. In the single-band approximation this system is described by a Bose-Hubbard Hamiltonian. The system has a superfluid and a Mott insulating phase which can be either paramagnetic or ferromagnetic. We characterize the quantum phase transitions at unit filling by means of a density-matrix renormalization group technique, and compare the results with a mean-field approach and an effective spin Hamiltonian. The presence of the ferromagnetic Ising-like transition modifies the Mott lobes. In the Mott insulating region the system maps to the ferromagnetic spin-1/2 XXZ model in a transverse field and the numerical results compare very well with the analytical results obtained from the spin model. In the superfluid regime quantum fluctuations strongly modify the phase transition with respect to the well established mean-field three dimensional classical bifurcation. 
\end{abstract}


\maketitle

\section{Introduction}

Ultra-cold atoms in optical lattices have opened new possibilities to study quantum phase transitions~\cite{Sachdev} 
and to observe the effects of quantum fluctuations~\cite{bloch_review,cazalilla2011}.
Recent experimental advances have also paved the way to the investigation of 
quantum magnetism, notable examples being the demonstration of super-exchange interactions in bosonic gases~\cite{anderlini2007},
the time-evolution of spin impurities~\cite{cataniImp,BlochImp}, and the engineering of
Ising~\cite{greiner2011} and anisotropic exchange Hamiltonians~\cite{esslinger2013,blochXXZ}.
On the other hand cold atoms are also very suitable to study coherence phenomena related 
to the control of the coupling between internal levels of atomic species. 
One can obtain coherently coupled superfluids, which show many interesting features ranging from a classical bifurcation transition in 
internal Josephson effect~\cite{OberthalerIJ} to dimerization of half-vortices in rotating superfluids~\cite{Son,Ueda}. 

In this work we combine the two ingredients by studying a coherently coupled Bose gas trapped in a one-dimensional (1D) optical lattice at unit filling, which can be described by a coupled two-component Bose-Hubbard model with on-site interactions (see Eq.(\ref{EqHam})). 
In particular the relative strengths of the coherent coupling (or phase coupling) and the density couplings due to species-dependent two-body interactions drive the system into superfluid (SF) or Mott-insulating (MI), non-polarized/paramagnetic (NP) or polarized/ferromagnetic (FM) phases.
We characterize the phase diagram in detail by combining mean-field and density matrix renormalization group (DMRG) approaches \cite{white}, and by mapping to spin chain Hamiltonians. The interest in such a system is manyfold since it allows for the study of different topics such as: the role of quantum fluctuations due to confinement and interaction in the NP-FM bifurcation in the superfluid regime; the change of the lobes in the SF-MI transition, which in 1D (at constant integer density) is of the Berezinskii-Kosterlitz-Thouless (BKT) type~\cite{BHmodelFisher,MonienDMRG,white2}; the Ising-like ferromagnetic transition in the MI phase; and the possible simulation of a ferromagnetic XXZ chain in a transverse field.
Moreover, the model Hamiltonian we use is relevant for ladder chain models in presence of a density-density interaction between the particles on different chains 
(see~\cite{Orignac1998,Nonna}, where the incommensurate filling case is studied), which has not been  as much studied as the case of non-interacting chains (see, e.g., \cite{Danshita2007} and references therein).

In systems of hard-core bosons or fermions with nearest-neighbor intra-species and on-site intra-species interactions, the NP-FM transition has been studied  for the density (charge) gapless phase \cite{Takayoshi}. 
Interestingly it has been shown that the transition belongs to the Ising in transverse field universality class. We find that the same holds for our model, but in MI, i.e., density (charge) gapped phase. By means of our accurate numerical tools we give an explicit expression for the phase transition point. Moreover we characterize completely the various phases, and find, e.g., as mentioned above, that the NP-FM transition affects the Mott lobes. We also determine the behavior of the transverse and longitudinal spin correlation functions across the phase transition. The latter quantities can be directly measured in cold gases experiments \cite{Greif, Hung}.
\section{Coherently Coupled Bose-Hubbard Model}

We consider a Bose gas at unit filling confined in a 1D geometry with two hyperfine levels that are coherently coupled. The atoms feel a deep optical lattice of  number of sites $L$ which is the same for the two internal levels. The system can be described by a two-component single-band Bose-Hubbard Hamiltonian with a static linear coupling between the two species:
\begin{align}
 H&= \sum_{i} \Big[\sum_\sigma  \frac{U}{2}\hat{n}_{i\sigma}(\hat{n}_{i\sigma}-1) +  U_{ab}\hat{n}_{ia}\hat{n}_{ib} \Big]+ \nonumber \\
   & + J_\Omega \sum_{i}( \hat{a}_i^\dag \hat{b}_i + \hat{a}_i\hat{b}_i^\dag)  -\frac{J}{2}\sum_{<ij>} (\hat{a}_{i}^\dag \hat{a}_{j}+\hat{b}_{i}^\dag \hat{b}_{j}+H.c.)\label{EqHam}
\end{align}
where $\sigma=a,b$ is the index distinguishing the two (pseudo-spin) internal levels, $\hat{a}_i$, $\hat{b}_i$ are the corresponding annihilation operators on the lattice site $i$ and  $\hat{n}_{i\sigma}$ is the number operator. 
The interaction terms $U$ and $U_{ab}$ represent on-site intra- and inter-species two-body interactions, respectively, while $J_\Omega$ is the strength of the conversion from one internal level to the other. Finally, the hopping with strength $J$, limited to nearest neighbours ($<\!\!ij\!\!>$), represents the kinetic energy in the lattice. In this work we restrict for the sake of clarity to equal intra-species interactions and equal hopping for both components. Equal hopping is also the typical situation in ultra-cold gases experiments.   

The presence of a static linear coupling $J_\Omega$, as the one employed in \cite{matthews,williams,OberthalerIJ, beattie, nicklas}, makes the system very different from the much studied Bose-Bose mixtures~\cite{Altman2003,DemmlerLukin,Kuklov2003,Ozaki2012,das} or from schemes where $J_\Omega$ is time dependent ~\cite{hazard1,hazard2,sun}. Shortly in the two component case one has two $U(1)$ symmetries (related to the conservation of the atom number in each species, being $J_\Omega=0$, and broken in the SF regime)  and when the interspecies interaction fulfills $U_{ab}> U$ the mixture phase separates  \cite{das}. In the presence of the interchange term only one $U(1)$ symmetry is left, the system is always miscible and if $U_{ab}$ is large enough a $\mathbb{Z}_2$ symmetry is broken allowing for a second order phase transition which brings the system to a FM state. Notice also that the miscible-immiscible transition for mixtures (phase separation) is of the first order kind.

\section{ Mott-Superfluid Phase Transition}

The Mott-superfluid transition is related to the breaking of the $U(1)$ symmetry, which leads to the emergence of a global phase, and thus to quasi-condensation in 1D.
In the absence of hopping, $J=0$, the ground state of Hamiltonian Eq.~(\ref{EqHam}) is $\ket{0}=\prod_i c_i^\dag\ket{\text{vac}} $, where $\ket{\text{vac}}$ is the vacuum of particles and we have introduced the operators $\hat{c}^\dag_i=(\hat{a}^\dag_i-\hat{b}^\dag_i)/\sqrt{2}$ creating a particle in site $i$ in the anti-symmetric state of the internal levels $a$ and $b$ (dressed state). Notice that if $J_\Omega$ were not real (or positive), a different relative phase would appear between $\hat{a}^\dag$ and $\hat{b}^\dag$ in the definition of $\hat{c}^\dag$ which would not affect the properties of the system.

\begin{figure}[h]
 \epsfig{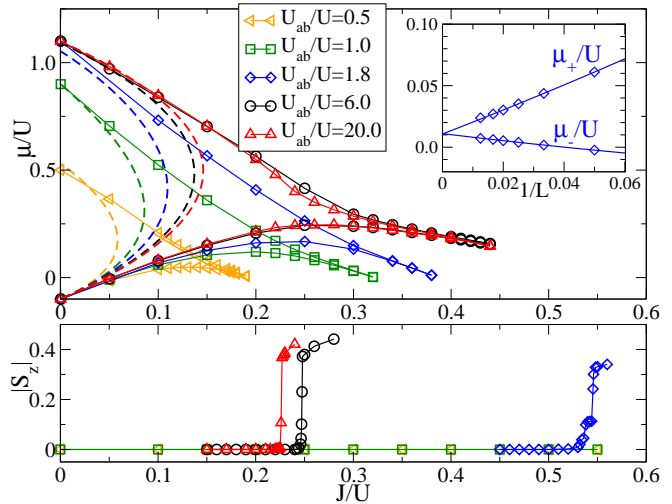}
 \caption{Top panel: MI-SF phase transition predicted by the mean-field approach (dashed lines) and DMRG (symbols). In the inset we show a typical finite size scaling of $\mu^+$ and $\mu^-$ in the superfluid regime for $U_{ab}/U=1.8$. We characterize a charge gapless phase, i.e. superfluidity, by $\mu^+-\mu^-=0$ in the thermodynamic limit. Bottom panel: associated NP-FM transition calculated with DMRG. All curves correspond to $J_\Omega/U=0.1$ and solid lines are drawn as a guide to the eye.}\label{FigLobes}
\end{figure}

In the presence of hopping the system undergoes a phase transition between a Mott insulating and a superfluid phase. 
It is customary to depict the phase diagram of the system as a function of its chemical potential $\mu$ and the tunnelling energy $J$. This leads to a lobe structure with fixed  filling within the Mott lobes. 
Examples of phase diagrams of Hamiltonian (~\ref{EqHam}) for $n=1$ are plotted in Fig.~\ref{FigLobes} (top panel), both within mean-field approximation and the exact DMRG result (see text below).

\subsection{Mean Field Mott-Superfluid Phase Transition}

In order to get an insight into the way the different parameters of the model enter in the SF-MI phase transition, we apply a mean-field theory \cite{Sachdev} to the grandcanonical Hamiltonian $H-\mu\sum_{i\sigma}\hat{n}_{i\sigma}$. 
At $J=0$, the borders of the Mott lobes are easily determined by requiring unit filling factor. The chemical potential must satisfy the conditions $-J_\Omega <\mu $ and $\mu < J_\Omega +(U+U_{ab})/2-\sqrt{16J_\Omega^2+(U-U_{ab})^2}/2 $. 
For $J\neq0$, second-order perturbation theory predicts that the border between the MI and the SF region is given by the condition 
\begin{align}
  & \frac{1}{zJ}=\frac{1}{\mu+J_\Omega} + \frac{-2\mu+6J_\Omega+U+U_{ab}}{\left(-\mu+J_\Omega+\frac{U+U_{ab}}{2}\right)^2-4J_\Omega^2-\left(\frac{U-U_{ab}}{2}\right)^2},\label{EqMottMF}
\end{align}
where the coordination number is $z=2$ in 1D.  Notice that in the SU(2) symmetric case for the interaction, $U_{ab}=U$, the single component result is recovered provided the chemical potential is rescaled to $\tilde{\mu}=\mu-J_\Omega$. When the hopping strength $J$ becomes larger than that given by Eq.~(\ref{EqMottMF}) the system enters the SF phase and develops a nonzero order parameter given by $\psi_-=(\psi,-\psi)^T/\sqrt{2}$, with $\psi=\langle a\rangle=\langle b\rangle$. Since quantum fluctuations are neglected the MI phase is described by the state $\ket{0}$ introduced above. Therefore, the system could support a polarized state only in the SF regime provided $U_{ab}$ was large enough, in analogy to coupled condensates (see, e.g., the experiment reported in \cite{OberthalerIJ} and references therein). 

The structure of the mean-field Mott lobes given by Eq.~(\ref{EqMottMF}) is shown as dashed lines in Fig.~\ref{FigLobes} for different values of $U_{ab}$. 
There are a number of features in the structure of the lobes to be noticed: the lower border equals $-J_\Omega/U$ for all values of $U_{ab}/U$ and the upper border converges at $1+J_\Omega/U$ for $U_{ab}>U$;  as $U_{ab}/U$ is increased, the lobes saturate at a maximum value of $J/U$, a feature that also takes place in mixtures. 
Moreover at fixed $U$ one has, as expected by the change in the compressibility, that for $U_{ab}<U$ the insulating region is smaller than in the single component case, while for $U_{ab}>U$ the insulating region is enlarged. 

With respect to quantum systems in higher dimensions \cite{Zwerger}, in 1D the role of quantum fluctuations can bring relevant beyond mean-field effects \cite{cazalilla2011}. These are usually not properly captured in semi-classical approaches, such as the mean-field, but can be accounted for in quasi-exact methods such as DMRG (see next paragraph).

\subsection{DMRG Mott-Superfluid Phase Transition}

In order to check the previous analysis and to get quantitative results we use DMRG technique \cite{white} to determine the properties of the ground state of Eq.~(\ref{EqHam}). This method has already proven to give strong beyond mean-field effects in the context of the single-species Bose-Hubbard model \cite{MonienDMRG,white2}. All the numerical results are obtained at unit filling. 
The Mott lobes calculated with DMRG  are shown as symbols in Fig.~\ref{FigLobes} (top panel). 

As expected, we find that the Mott-superfluid transition takes place at values of $J/U$ much higher than predicted by mean field (dashed lines), and that the lobes have the reentrant shape characteristic of the 1D Bose-Hubbard Hamiltonian~\cite{MonienDMRG,white2}.
We determine the transition points by the closure of the so-called density (or charge) gap for different system sizes and then performing finite size scaling as we report in the inset of Fig. 1. The density gap, $\mu=\mu^+-\mu^-$, for a system with N particles with energy $E(N)$ is defined by the difference in energy in adding, $\mu^+ = E(N+1)-E(N)$, or removing, $\mu^-= E(N)-E(N-1)$, a particle.
While such a method works very well for incommensurate transition points it is known to be less accurate for determining the commensurate-commensurate transition. The latter belongs indeed to the BKT universality class with an exponentially small gap closure~\cite{white2}. However the use of the gap closure is enough 
for the purpose of the present work.
We have indeed checked that our transition points for $U=U_{ab}$ (equivalent to the single component case) are in very good agreement with the ones obtained by calculating the central charge as in \cite{kollath}.

\section{Para-/Ferro-Magnetic Phase Transition}

In addition to the Mott-SF transition, Hamiltonian~(\ref{EqHam}) allows for states breaking a $\mathbb{Z}_2$ symmetry, creating a finite polarization $S_z=(N_a-N_b)/2N$, with $N_\sigma$ the number of atoms in state $\sigma=a,b$. 

\subsection{Global Magnetization}

In order to study the breaking of the $\mathbb{Z}_2$ symmetry in Hamiltonian~(\ref{EqHam}) we first of all determine the global polarization (or magnetization), $S_z$.
In our numerical simulations this requires special attention, especially in the superfluid phase, since a sufficiently large size of the Hilbert space has to be taken. That is, we need to consider an on-site basis  containing the states corresponding to a number of bosons up to $n_{max}$, to allow the fluctuations of $a$ and $b$ to explore the relevant configurations and thus to drive the phase transition.
We obtain convergence of the results for open boundary conditions using $n_{max}=6$, keeping up to 512 DMRG states and 6 sweeps \cite{white}, getting a truncation error lower that $10^{-8}$. Unless otherwise stated we show the results for a chain with $L=N=80$ \cite{commentLuca}.

The results for the absolute value of the polarization \cite{comment2} as a function of $J/U$ are reported in the bottom panel of Fig.~\ref{FigLobes}.

In the SF phase (corresponding to $U_{ab}/U=1.8$) the system shows strong quantum fluctuations. Indeed, the NP-FM transition has been studied in the continuum and within the Gross-Pitaevskii framework (for a recent discussion see, e.g, \cite{Abad2013} and references therein), and has been seen to take place for $U_{ab}-U= 2J_\Omega/n$, with $n=1$ the total density of the system.
Moreover the critical exponent of the magnetisation is in this case the expected mean-field value $\beta=1/2$.
In the lattice, instead, the transition occurs for an inter-species interaction larger (but still of the same order) than the one predicted for a mean-field coherent state, i.e. $U_{ab}/U=1.2$, and the magnetization does not follow the classical bifurcation law. 
Notice that the magnetization behavior for $U_{ab}/U=1.8$ is not properly described by mean-field or strong coupling analyses. This makes it very challenging to explain the peculiar increase of $|S^z|$.
%Furthermore the system behavior for such a value of $U_{ab}$ is not properly described neither by the MF nor by strong coupling analisis. It makes this regime very challenging and it could explain the peculiar growing of $|S^z|$. 

In the Mott phase, where double occupancy is strongly suppressed, the inter-species interaction has to be much stronger, e.g. $U_{ab}/U=6$ and $U_{ab}/U=20$, to drive the phase transition. For increasing values of $U_{ab}$ the transition point is seen to approach a limiting value of $J$  corresponding to the value given by the ITF mapping discussed above.

Moreover, it can be noticed from Fig.~\ref{FigLobes} that once the magnetic phase transition has taken place inside the lobe (see for instance the case $U_{ab}/U=20$), the latter shrinks slightly, indicating that the SF phase is more favorable than the MI for the polarized system. 
Also, in this case the Mott insulating lobes do no longer strongly depend on the value of $U_{ab}$, since in the ferromagnetic phase this interaction is less effective. This saturation of the Mott lobes for $U_{ab}$ large has a completely different meaning from the saturation found in the mean-field analysis.

\subsection{Strong Coupling Regime}

When the system becomes strongly interacting the fluctuations of the number of atoms in each site are weaker and therefore the effect of the two-body interaction is reduced, making the polarized state less favorable. In particular in the deep MI phase ($J\ll U$, $U_{ab}$) the single particle tunneling is suppressed and exchange of atoms is the dominant process. In this case the coherently coupled Bose-Hubbard model Eq.~(\ref{EqHam}) can be mapped into a spin chain model (see, e.g., \cite{Kuklov2003,DemmlerLukin}). The effective spin Hamiltonian is the so-called spin-$1/2$ XXZ model in a transverse field (see, e.g., \cite{Dmitriev2002}), which reads 
\begin{equation}
H_{XXZ}=-t\sum_i (\hat{S}_i^x\hat{S}_{i+1}^x+\hat{S}_i^y\hat{S}_{i+1}^y+\Delta \hat{S}_i^z\hat{S}_{i+1}^z)+2J_\Omega\sum_i \hat{S}_i^x,
\label{eq:XXZ}
\end{equation}
where  $\hat{S}_i^z=(\hat{n}_{i a}-\hat{n}_{ib})/2$, $\hat{S}_i^x=(\hat{a}_i^\dag \hat{b}_i + \hat{a}_i\hat{b}_i^\dag)/2$, 
$\hat{S}_i^y=-i(\hat{a}_i^\dag \hat{b}_i - \hat{a}_i\hat{b}_i^\dag)/2$, $t=4J^2/U_{ab}$ and  $\Delta=2U_{ab}/U-1$ is the anisotropy. 
Since we are considering repulsive on-site interactions we are restricted to $-1<\Delta<+\infty$. In such parameter range the spin 
model Eq.~(\ref{eq:XXZ}) exhibits only two phases, a paramagnetic phase with magnetization along the $x$-axis and an Ising ferromagnetic 
phase along the $z$-axis. For $J_\Omega=0 $ the model is exactly solvable and the transition occurs at $\Delta=1$, i.e., $U_{ab}=U$. 
For $J_\Omega\neq 0$ the transition is shifted to larger values of $U_{ab}/U$. On the other hand for $U_{ab}\rightarrow\infty$ the 
Hamiltonian reduces to the Ising model in a transverse field (ITF) which is also exactly solvable and predicts 
a transition at $t\Delta=4J_\Omega$, i.e., for $2J^2=UJ_\Omega$, with a critical exponent $\beta=1/8$. 
The mapping to ITF tells us that even in the infinite inter-species interaction case one always needs a minimum tunneling to observe the ferromagnetic transition. As we will explain in detail in the next paragraphs we find that the magnetic phase transition in the MI phase belongs indeed to the ITF universality class, in analogy with the results obtained in \cite{Takayoshi}.

\begin{figure}[h]
 \epsfig{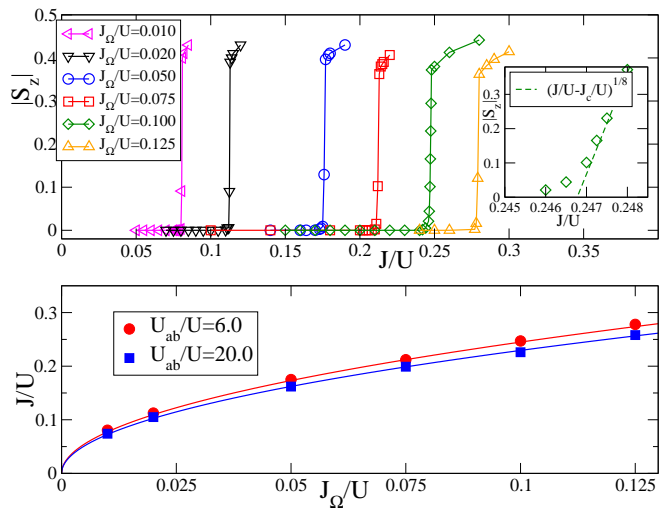}
 \caption{ Top panel: NP-FM transition in the MI phase for different values of the linear coupling $J_\Omega/U$, for $U_{ab}/U=6$. Inset: comparison between numerical results and critical exponent $1/8$ of ITF for $J_\Omega=0.1$. Bottom panel: NP-FM transition point calculated with DMRG (symbols) and using expression $J^2(1/U-1/U_{ab})=J_\Omega/2$ (solid lines, see text and Eq.~(\ref{eq:PTpoint}) for more details), for two values of $U_{ab}/U$ in the MI phase.}\label{PTvsJO}
\end{figure}

Let us better characterize the FM transition in the MI regime by changing $J_\Omega$, as reported in the top panel of Fig.~\ref{PTvsJO}, which shows the DMRG results. As described above in the Mott phase for $J_\Omega\rightarrow0$  the system is equivalent to the XXZ model, which gives the FM transition at $U_{ab}/U=1$.  For $J_\Omega\neq 0$ the transition is shifted to larger values of $U_{ab}/U$. One can obtain an approximation to the critical condition by noticing that Hamiltonian Eq.~(\ref{eq:XXZ}) can be rewritten as a Heisenberg exchange term, $\sum \vec{S}_i\cdot \vec{S}_{i+1}$, plus an ITF term. 
Neglecting the effect of the Heisenberg term (valid for $U_{ab}>U$), the phase transition is driven by the ITF and it takes place at 
\begin{equation}
t(\Delta-1)=8J^2(1/U-1/U_{ab})=4 J_\Omega.
\label{eq:PTpoint}
\end{equation}
The accuracy of this expression with respect to the numerical solution of Hamiltonian Eq.~(\ref{EqHam}) is shown in the bottom panel of Fig.~\ref{PTvsJO}, where it is seen to be very good for a range of values of $J_\Omega$. Moreover in the inset of Fig.~\ref{PTvsJO} it is possible to notice that the critical exponent $\beta=1/8$ of ITF is in good agreement with our numerical data. Such results justify the use of the spin model to address the magnetic properties of Bose gases in optical lattices also for not too small values of $J/U$.

\subsection{Spin-Spin Correlation Functions}

While $S_z$ is the global order parameter, we characterize the NP and FM phases, and in particular the NP-FM transition, also by determining the behavior of the correlation functions around the phase transition point. We study the longitudinal and the transverse spin-spin correlation functions  $C_s(i)=\langle \hat{S}^s_j\hat{S}^s_{j+i}\rangle$  with $s=z,x$ respectively.
In order to drop boundary effects we exclude the more external sites and  evaluate the correlation functions only in the central region of the system (in particular we take $j=15$).

\begin{figure}
 \epsfig{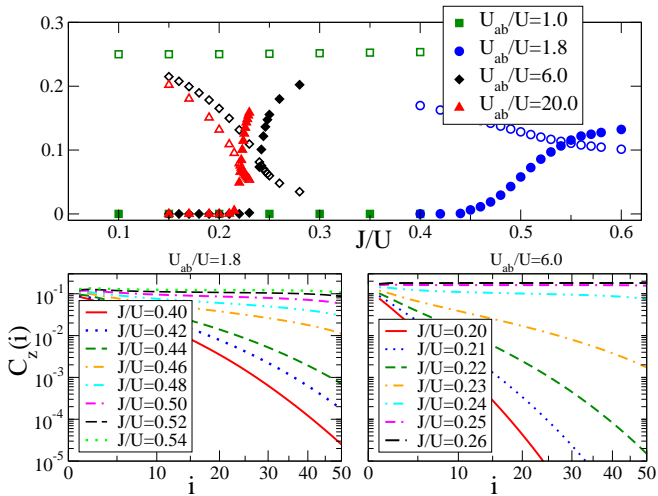}
 \caption{Top panel: Behavior of $C_x(50)$ (open symbols) and $C_z(50)$ (filled symbols) across the MI-SF transition, for $J_\Omega/U=0.1$ (see Fig.~\ref{FigLobes}). Bottom panels: Long-range behavior of $C_z(i)$ in the SF (left panel) and MI (right panel) phases close to the phase transition.}\label{fig:core}
\end{figure}

To have an idea of how the large distance behavior of the correlation functions changes along the transition, we plot in the top panel of Fig.~\ref{fig:core} the correlation functions for a separation $i=50$ as a function of $J/U$.  
The paramagnetic phase is dominated by transverse spin correlations since in this regime $J_{\Omega}$ is the most important term, while in the ferromagnetic phase the longitudinal correlations become dominant. Notice that the magnetic transition (see Fig.~\ref{FigLobes}) seems to be well described by the crossing point between the long-range values of $C_x$ and $C_z$. 

The longitudinal correlation function across the NP-FM transition is shown in the lower panels of  Fig.~\ref{fig:core} in the superfluid ($U_{ab}/U=1.8$, left panel) and in the MI phase ($U_{ab}/U=6$, right panel). The behavior of $C_z$ changes from an exponential decay in the paramagnetic phase to long-range order in the FM phase showing a clear second order phase transition. The critical point is in good agreement with the one obtained with $S_z$ (Fig.~\ref{FigLobes}). Notice that in the SF phase the system polarizes more ``slowly'' than in the insulating case due to strong fluctuations, which explains the larger region of intermediate decays. 

\section{Conclusion and Perspectives}

Let us briefly comment here on the experimental realization of the model Hamiltonian Eq. (\ref{EqHam}) with cold-gases. Even if some relevant ingredients are already available within the present technology some challenging achievements are missing. The two species Bose-Hubbard models have been realized and their mapping to a spin chain tested, see, e.g., \cite{blochXXZ}. Adding a static Rabi coupling is not an issue. At the same time, in fermionic systems, temperatures of the same order of the spin exchange have been reached \cite{esslinger2013}. In current experiments, where $Rb$ atoms are used, the most difficult and not yet achieved ingredient is to have very different intra- and inter-species interaction to address the ferromagnetic transition in the MI phase. 
A very helpful tool in this direction is the recent possibility, explored in Esslinger's group \cite{EsslingerStateDepLattice}, of creating state-dependent lattices for essentially any atomic species.  At the same time spin-selective microwave fields could allow for the exploration of resonances in non-standard collision channels \cite{dalibard}. It would open the way towards the achievement of a large range $U_{ab}/U$ values.

In conclusion, the system we have studied, described by Eq.~(\ref{EqHam}), constitutes a quite unexplored system in the family of Bose-Hubbard Hamiltonians (see, e.g., also \cite{Orignac1998,Takayoshi}). It is fundamentally different from Bose-Bose mixtures and in a way a generalization of two-leg chains. The system shows two quantum phase transitions: superfluid to Mott insulator transition -- which is of the Berezinskii-Kosterlitz-Thouless kind at fixed integer density -- and a paramagnetic/non-polarized to ferromagnetic/polarized transition. We show that the latter transition changes the structure of the Mott lobes. In the Mott regime the transition is well described in terms of a quantum XXZ model in a transverse field. In the SF regime due to quantum fluctuations strong corrections to the mean-field coherent results are present.  While we focused on the unit filling factor case, at low filling factor, the system is also interesting, especially considering that its experimental realization should be feasible within current technology as shown in \cite{blochXXZ}. Indeed in the small $J/U$ case both species $a$ and $b$ have a fermionic (Tonks-Girardeau regime) equation of state \cite{LiebLiniger}. Therefore one has the possibility of studying the fate of itinerant ferromagnetism in one dimension in analogy to the recent analysis in  \cite{HoStoner1D} with the inclusion of the linear interspecies coupling $J_\Omega$. Another interesting aspect to study is the dynamics of the system. The latter has been studied in some detail for the homogeneous weakly interacting case. In the presence of a lattice it would be interesting, e.g., to study the quenching across the ferromagnetic transition \cite{LeeGao,SabatiniZurek,DalllaTorreDemler} or how $J_\Omega$ would modify the domain wall dynamics (see, e.g., \cite{McCullochMelting}) or the quenching across the ferromagnetic transition.

{\it Acknowledgement.} Useful discussions with and G. Ferrari, Yan-Hua Hou and Tommaso Roscilde are acknowledged. 
This work has been supported by ERC through the QGBE grant and by Provincia
Autonoma di Trento. L.B. acknowledges support by Cariparo Foundation (Eccellenza grant 11/12)
and the CNR-INO BEC Center in Trento for CPU time. A.R. acknowledges support from the Alexander von Humboldt foundation. 
M. A. acknowledges support from the Okinawa Institute of Science and Technology Graduate University during the final stages of the work.

During the review process of the current manuscript, the Mott regime has been studied \cite{zhan} obtaining results in agreement with ours.

\end{document}